\documentclass[aps,twocolumn,amsmath,amssymb, superscriptaddress, nofootinbib]{revtex4}
\usepackage[pdftex]{graphicx}% Include figure files
 \usepackage{amsmath}
\usepackage[T1]{fontenc}
\usepackage{amssymb}
\usepackage{amsfonts}
\usepackage{bm}
 \usepackage{amsmath} 
  \usepackage{flushend}
\usepackage{color}
\usepackage{lipsum}
\usepackage{amsfonts} 
\usepackage{amssymb, mathrsfs}
\usepackage{braket}
\usepackage{graphicx} 
\usepackage{subfigure}
\usepackage{bbm}
\def\beq{\begin{equation}}
\def\eeq{\end{equation}}
\def\bsp{\begin{split}}
\def\esp{\end{split}}
\def\bea{\begin{eqnarray}}
\def\eea{\end{eqnarray}}
\def\ba{\begin{array}}
\def\ea{\end{array}}

\def\dg{\dagger}

\def\lb{\left(}
\def\rb{\right)}

\def\l.{\left.}
\def\r.{\right.}

\def\ra{\rangle}
\def\la{\langle}

\def\bo{{\vec k}}

\begin{document}

\date{\today}
\title{Strain-Induced Topological Magnon  Phase Transitions:\\ Applications to Kagome-Lattice Ferromagnets}
%\email{solomon@aims.ac.za}
\author{S. A. Owerre}
\affiliation{Perimeter Institute for Theoretical Physics, 31 Caroline St. N., Waterloo, Ontario N2L 2Y5, Canada.}

\begin{abstract}
A common feature of topological insulators is that they are characterized by topologically invariant quantity such as the Chern number and the $\mathbb{Z}_2$ index. This quantity distinguishes a nontrivial topological system from a trivial one. A topological phase transition may occur when there are two topologically distinct phases, and it is  usually defined by a gap closing point where the topologically invariant quantity is ill-defined. In this paper, we  show that the magnon bands in the  strained (distorted) kagome-lattice ferromagnets realize  an example of a topological magnon phase transition in the realistic parameter regime of the system.  When spin-orbit coupling (SOC) is neglected (i.e. no Dzyaloshinskii-Moriya interaction),  we show that all three magnon branches are dispersive with no flat band, and there exists a critical point where  tilted Dirac and semi-Dirac point  coexist  in the magnon spectra.  The critical point separates two gapless magnon phases as opposed to the usual phase transition.  Upon the inclusion of SOC,  we realize  a topological magnon phase transition point at the critical strain $\delta_c=\frac{1}{2}\big[ 1-(D/J)^2\big]$, where  $D$ and $J$ denote the perturbative SOC and the Heisenberg spin exchange interaction respectively. It separates two distinct topological magnon phases  with different Chern numbers for $\delta<\delta_c$ and for $\delta>\delta_c$.  The associated anomalous thermal Hall conductivity  develops an abrupt change at $\delta_c$, due to the divergence of the  Berry curvature in momentum space. The proposed topological magnon phase transition is experimentally feasible  by applying external perturbations such as uniaxial strain or pressure.
\end{abstract}
\maketitle

%In two-dimensional (2D)  distorted honeycomb lattice (or deformed graphene), a semi-Dirac point can be obtained by emerging  two Dirac points. It appears as a single entity at the phase transition   between a gapless and trivial phase. In the presence of spin-orbit coupling (SOC), the semi-Dirac point remains gapless, but appears at the topological phase transition between a topological and trivial insulator.  The magnon bands in the distorted honeycomb-lattice ferromagnets mimic exactly the same electronic band theory.

\section{Introduction}

The two-dimensional (2D) graphene sheet with negligible SOC is the simplest example of Dirac points (DPs), where  two bands cross linearly  in momentum space \cite{cas}. They usually occur at the high symmetry points in the Brillouin zone (BZ), and protected by the coexistence of inversion and time-reversal symmetry.  In the  distorted graphene with unequal hopping integrals,  a semi-Dirac point (SDP)  with linear band crossing along one momentum direction and quadratic band crossing along the perpendicular momentum direction, can be realized  by emerging two DPs \cite{mon, mon1, mon2,mon3,mon4,mon5}. It defines a phase transition between a gapless and a gapped trivial phase. When SOC is not neglected, the SDP transforms to a gapless DP at the topological phase transition  between  topological and  trivial insulator \cite{lang,mur}. As the notion of band theory is independent of the quasiparticle excitations, the magnon bands in the 2D (distorted) honeycomb ferromagnets \cite{mag,owe1,boy, yago,per} directly mimic that of (distorted) graphene. To our knowledge, however, the coexistence of (tilted) DPs and SDPs in a single system has not been studied. 

% Recently, tilted (type-II) Dirac cones have also emerged as a new focus in Dirac materials \cite{tt1,tt2,tt3}.

Furthermore, the ideal insulating kagome-lattice ferromagnets usually allow a SOC  in the form of the  Dzyaloshinskii-Moriya (DM) interaction \cite{dm, dm2}, due to lack of an inversion center. The DM interaction generally leads to   magnonic topological insulators (mTIs) \cite{mp1,mp2,mp3, mp4}, with similar properties (such as the appearance of chiral edge modes and  Chern numbers) to electronic topological insulators (TIs) \cite{top1, top2, top3, top4}. However, the experimental observation of mTIs remains elusive, having been observed only  in the quasi-2D  kagome ferromagnet Cu(1-3, bdc)  \cite{alex6}, which exhibits  a nonzero anomalous thermal Hall effect \cite{alex6a}.  The elusiveness of a direct  experimental observation of mTIs  is in part  due to the fact  that real  kagome ferromagnetic materials may have very weak and negligible  DM interaction as recent experiments have shown in the spin-$1/2$ kagome ferromagnetic mineral haydeeite, $\alpha$-MgCu$_3$(OD)$_6$Cl$_2$ \cite{bol}. Therefore, DPs may be intrinsic to such kagome materials with negligible effect of the DM interaction.  Moreover, most  real kagome ferromagnetic materials are imperfect and have intrinsic structural distortion  that deviates from the ideal structure. Besides, structural distortion  can also be achieved experimentally be applying external perturbation such as strain or pressure \cite{st1,st2,st3,st4}.   Hence, it is highly desirable to study how structural distortion affects the magnon physics in Dirac and topological magnetic materials.

In this paper, we present an exposition of the magnon physics in the insulating 2D strained (distorted) kagome ferromagnets.  We show that  when the DM interaction is negligible, there is no flat magnon band in the strained kagome ferromagnets, and the DPs do not appear at the high-symmetry points in the BZ. Interestingly, there exists a critical distortion point  where  titled DPs and SDPs coexist in the magnon spectra. The critical point separates two gapless magnon phases with DPs. However, when the DM interaction  is not neglected, there exists a topological phase transition point at $\delta_c$, separating two distinct  mTIs  with Chern numbers $(-1,1, 0)$ and   $(-1,0,1)$. The topological magnon phase transition point  generates  a divergent Berry curvature, which leads to an abrupt change in the associated anomalous thermal Hall conductivity. The proposed topological magnon phase transition appears in the realistic parameter regime in real materials, {\it i.e.}, $D<J$ \cite{alex6}, where $D$ is the perturbative DM anisotropy, and $J$ is the Heisenberg spin exchange interaction. Hence, we believe that the current topological magnon phase transition can be experimentally feasible in the kagome-lattice ferromagnets such as Cu(1-3, bdc)   with DM interaction $D/J=0.15$ \cite{alex6}, yielding a strained-induced topological magnon phase transition  at $\delta_c=0.48875$. This can be achieved by  external perturbations such as applied uniaxial strain or pressure.

\section{Strained Kagome Ferromagnets Without Dzyaloshinskii-Moriya interaction}
\subsection{Spin model}
\begin{figure}
\centering
\includegraphics[width=1\linewidth]{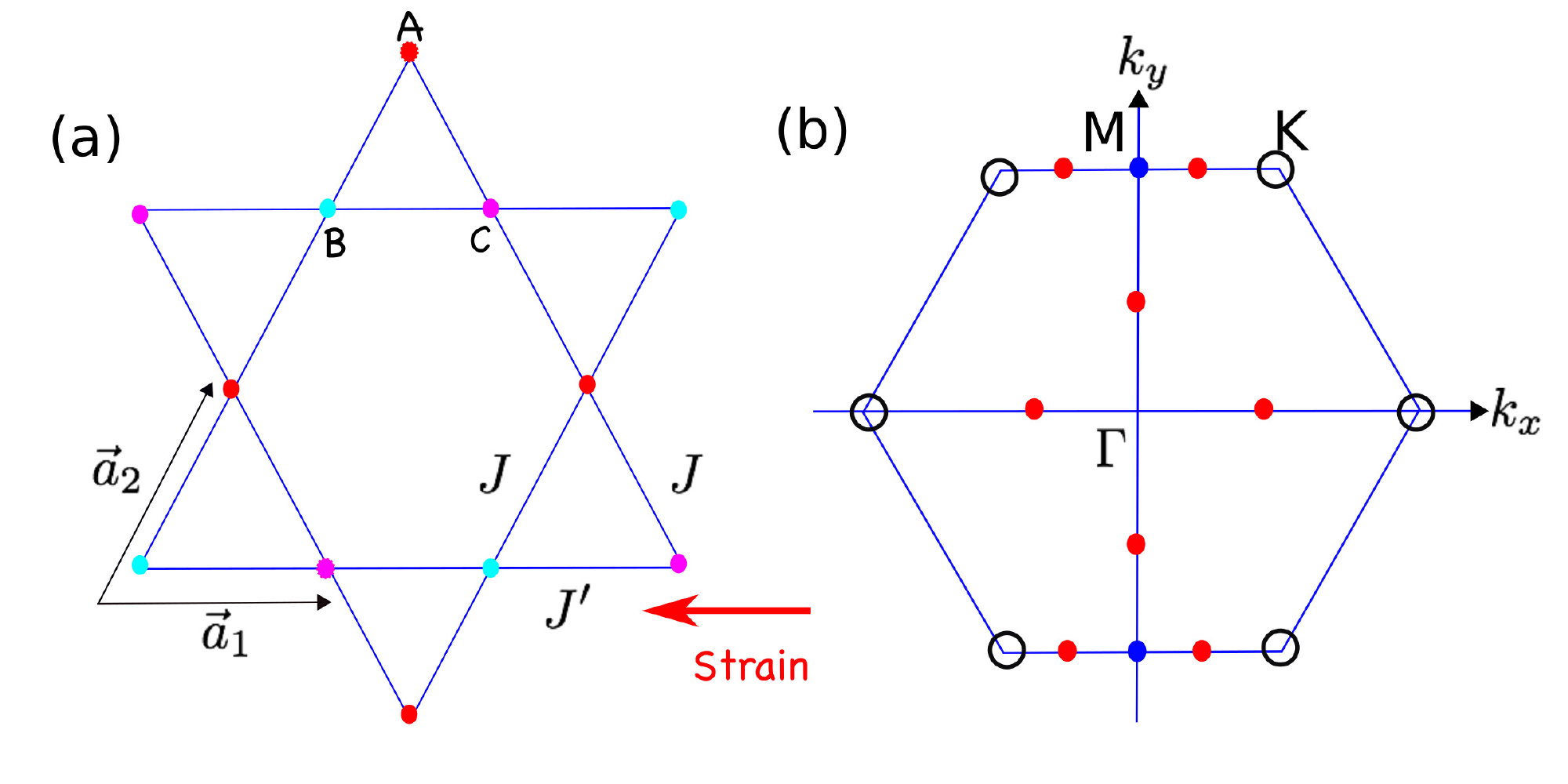}
\caption{Color online.  (a) Schematic of the kagome lattice with unaxial strain along the $x$-axis.   (b) The hexagonal BZ of the kagome lattice. Open black circles denote the locations of the Dirac magnon cones in the ideal lattice $J^\prime=J$. The red and blue dots denote the locations of the Dirac and semi-Dirac magnon cones in the strained lattice $J^\prime\neq J$ respectively. }
\label{lattice}
\end{figure}

We commence with the spin Hamiltonian for strained (distorted) kagome-lattice ferromagnets in the absence of the DM interaction, 
\begin{align}
\mathcal H_0&=-\sum_{ \la \ell \ell^\prime\ra } J_{\ell \ell^\prime}{\vec S}_{\ell}\cdot{\vec S}_{\ell^\prime}-B\sum_\ell S_\ell^z,
\label{spinh}
\end{align}
where ${\vec S}_{\ell}$ is the magnetic spin moment at site $\ell$, and the first summation runs over  nearest-neighbour (NN) spins.  $J_{\ell \ell^\prime}=J$ on the diagonal bonds and $J_{\ell \ell^\prime}=J^\prime=J\delta$ on the horizontal bonds, with $\delta \neq 1$  as shown in Fig.~\ref{lattice}(a). To model the effect of strain, we have made the assumption that only the interaction  along the $x$-axis changes, without lattice deformation\footnote{Alternatively, one could consider isotropic Heisenberg interactions with lattice deformation, {\it i.e.}, only the primitive lattice vectors change.}. The second term is an external magnetic field along the out-of-plane direction in units of $g\mu_B $. There are several limiting cases of this Hamiltonian.  First, $\delta=1$  corresponds to an ideal kagome-lattice ferromagnets. Second, $\delta= 0$ maps to a decorated  square lattice,  with additional sites at the midpoints of square lattice edges \cite{fa}. Third, $\delta\to \infty$ maps to a quasi-1D ferromagnetic spin chain. The point $\delta=0.5$ or $1/2$ is critical in the magnon spectra, and separates two gapless magnon phases as we will show below.  

%We note although the kagome lattice lacks inversion center, the perturbative DM interaction anisotropy can be negligible in  some materials, example includes  $\alpha$-MgCu$_3$(OD)$_6$Cl$_2$ \cite{bol}. Therefore, the ferromagnetic spin Hamiltonian \eqref{spinh} is applicable to real materials.  
  
\subsection{Bosonic model}
As the ground state of Eq.~\eqref{spinh} is a fully aligned ferromagnetic order, we describe the underlying magnetic excitations  by the Holstein Primakoff (HP)  transformation:  $S_{\ell}^{ z}= S-a_{\ell}^\dagger a_{\ell},~S_{\ell}^+\approx \sqrt{2S}a_{\ell}=(S_{\ell}^-)^\dg,$ where $a_{\ell}^\dagger (a_{\ell})$ are the bosonic creation (annihilation) operators, and  $S^\pm_{\ell}= S^x_{\ell} \pm i S^y_{\ell}$ denote the spin raising and lowering  operators. In the following, we set $B=0$  as it only rescales the lowest magnon band at the $\Gamma$-point.  The resulting noninteracting bosonic Hamiltonian is given by $\mathcal H_{\text{sw}}(\vec{k})=\Lambda_0{\rm I}_{3\times 3}-\Lambda(\vec{k})$, where $\Lambda_0=t\text{diag}(4,~2(1+\delta),~2(1+\delta))$
\begin{align}
\Lambda(\vec{k}) &=2t
\begin{pmatrix}
0& \cos k_2& \cos k_3\\
\cos k_2&0&\delta\cos k_1\\
\cos k_3&\delta\cos k_1&0
\end{pmatrix},
\label{honn}
\end{align}
 with   $t=JS$ and $k_i=\vec{k}\cdot\vec{a}_i$. The primitive vectors are  given by $\vec{a}_1=\hat x$, $\vec{a}_2=(\hat x,\sqrt{3}\hat y)/2$, and $\vec{a}_3=\vec{a}_2-\vec{a}_1$. 
\begin{figure*}
\centering
\includegraphics[width=.85\linewidth]{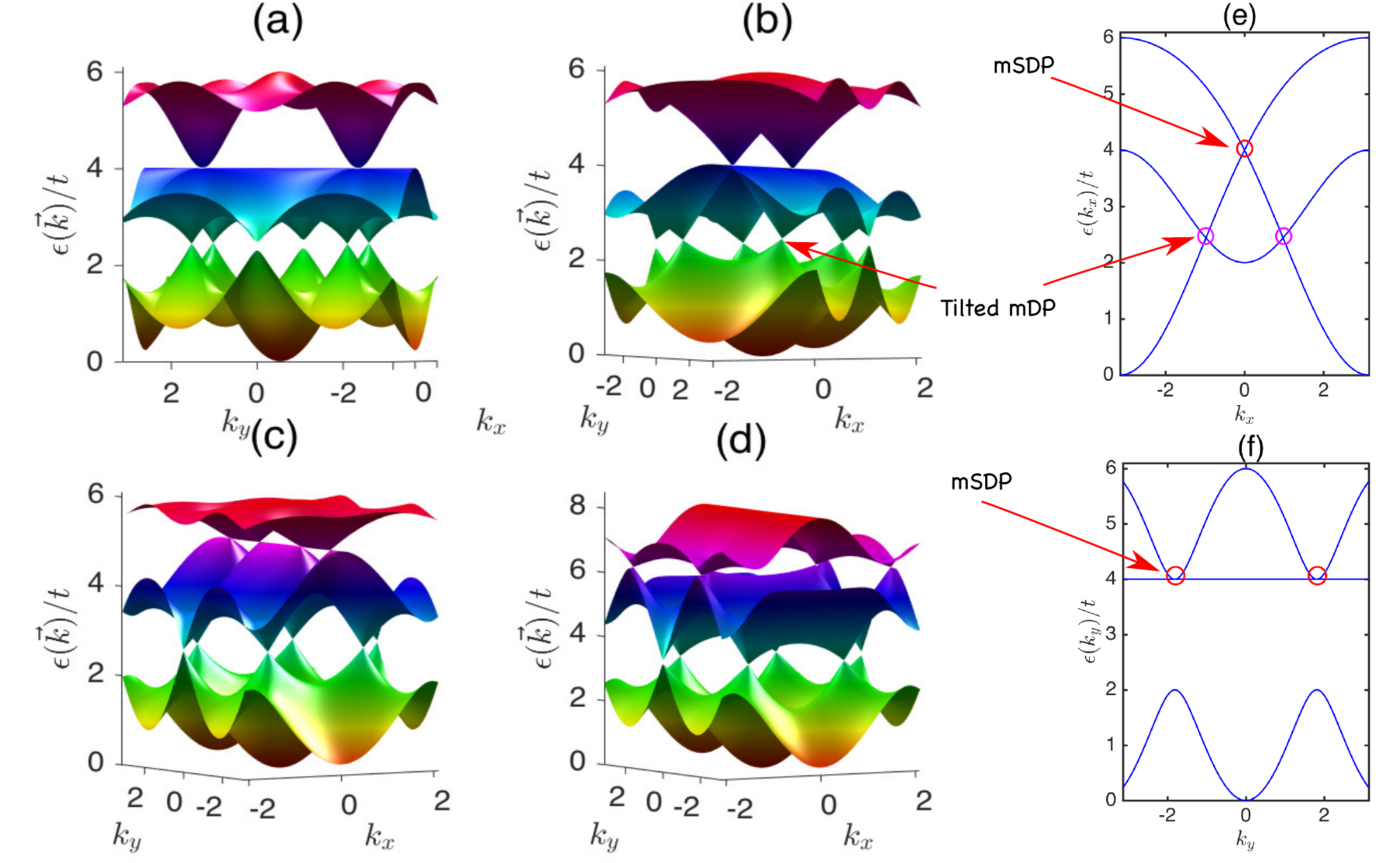}
\caption{Color online. (a,b) Coexistence of tilted Dirac  and semi-Dirac magnon cones in the distorted kagome ferromagnets $\delta=0.5$. (c,d) Dirac magnon cones in the distorted kagome ferromagnets for $\delta=0.75$ and $\delta=1.5$ respectively. (e, f) Coexistence of tilted mDP (pink circles) and mSDP (red circles) in the distorted kagome ferromagnets along (e) $k_x$ direction at $k_y=\pi/\sqrt{3}$ and along  (f) $k_y$ direction at $k_x=0$ for $\delta=0.5$. }
\label{band}
\end{figure*}
\begin{figure*}
\centering
\includegraphics[width=.85\linewidth]{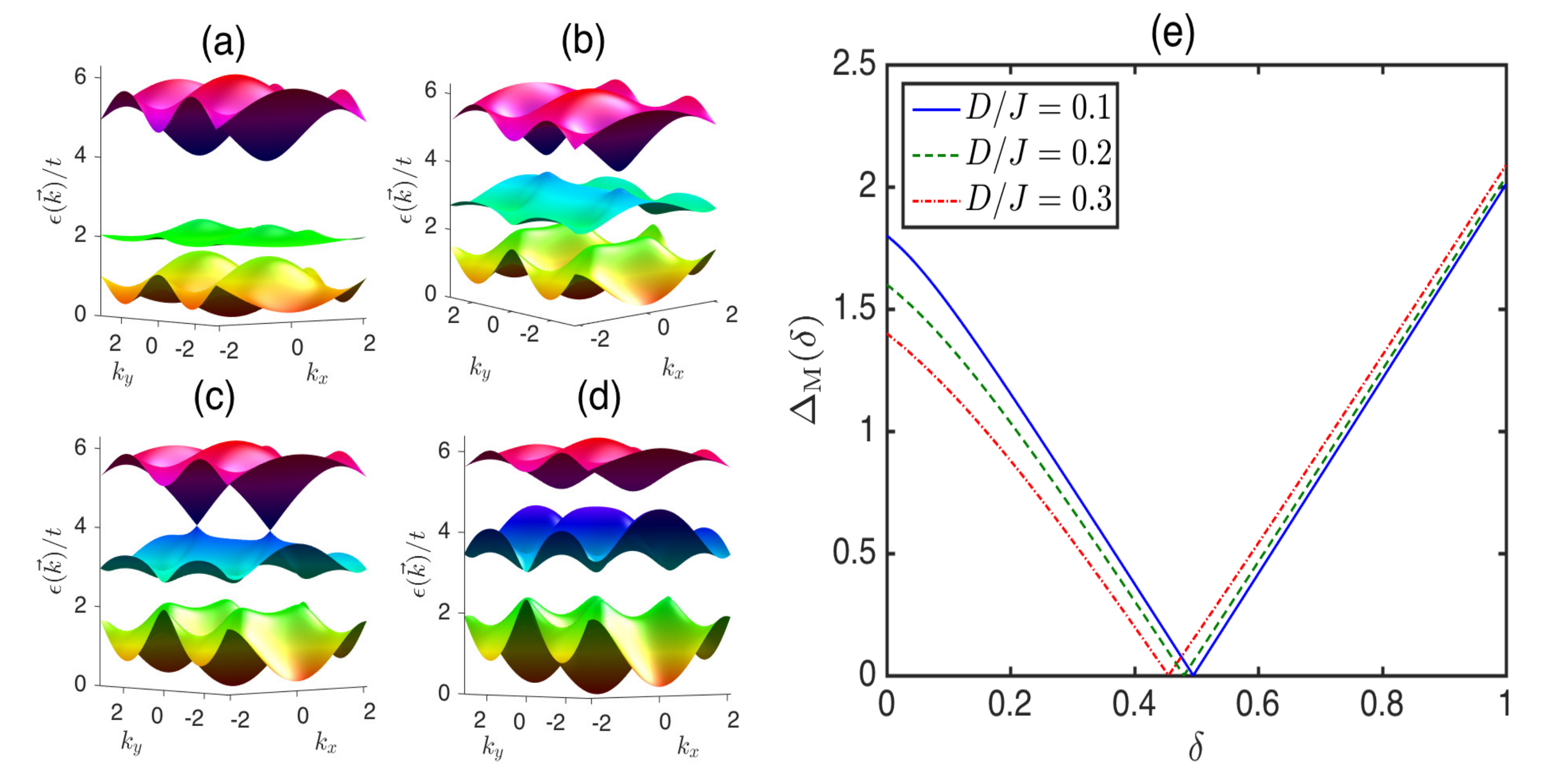}
\caption{Color online. (Left) Evolution of the massive Dirac magnons with nonzero DM interaction $D/J=0.2$. (a) $\delta=0$, (b) $\delta=0.35$, (c) $\delta=\delta_c=0.48$, (d) $\delta=0.75$. (Right, e) The gap between $\epsilon_{2,3}(\bo)$ magnon branches at the $\rm M$-point as a function of $\delta$ for different values of $D/J$. The gap vanishes at $\delta_c$ and separates two mTIs with different Chern numbers (see text). }
\label{banddm}
\end{figure*}
\subsection{Dirac and Semi-Dirac magnons}
For $\delta\neq 1$, the analytical  diagonalization of the Hamiltonian $\mathcal H_{\text{sw}}(\vec{k})$ is not feasible. Therefore we consider specific points in the BZ. Henceforth, we label the eigenvalues denoting the magnon bands as  $\epsilon_{\alpha}(\bo)$, where $\alpha=1,2,3$ denote the lowest, middle, and topmost magnon bands respectively.  At the  $\text{K}$-points (see Fig.~\ref{lattice}(b)), the   two magnon branches $\epsilon_{1,2}(\bo)$ are fully gapped out for $\delta\neq 1$ with the gap given by 
\begin{align}
\Delta_{\text{K}}(\delta)=\frac{t}{2}\lb \sqrt{12+3\delta(3\delta-4)}-(2+\delta)\rb.
\end{align}

In the ideal kagome-lattice ferromagnets, i.e., $\delta= 1$ the gap $\Delta_{\text{K}}(1)= 0$, which  leads to magnon Dirac points (mDPs) within the magnon branches $\epsilon_{1,2}(\bo)$  at the high-symmetry points of the BZ, as indicated by open circles in Fig.~\ref{lattice}(b). There is also a quadratic magnon band touching point at the $\Gamma$-point within the magnon band $\epsilon_{2}(\bo)$ and the flat magnon band $\epsilon_{3}(\bo)$ for $\delta= 1$.   In contrast,  for $\delta\neq 1$ there is no flat magnon band, and there are mDPs in all the three magnon branches $\epsilon_{1,2,3}(\bo)$, as well as magnon semi-Dirac points (mSDPs).  The mDPs occur away from the high-symmetry points in the BZ, whereas the mSDPs occur at the M-point as indicated by red and blue dots in Fig.~\ref{lattice}(b).

For $\delta\neq 1$, the  magnon branches $\epsilon_{1,2}(\bo)$ cross linearly (mDPs) at four points  as opposed to a total of six  for $\delta=1$. They are located at $\text{D}_1=(\pm k_x^{\text{D}_1},\pm\pi/\sqrt{3})$, where
\begin{align}
k_x^{\text{D}_1}=\arccos\lb-\frac{1+2\delta(1-\delta)-f(\delta)}{4\delta^2}\rb,
\end{align}
with $f(\delta)=\sqrt{1+4\delta\big[1+\delta\lb 2-\delta(2-\delta)\rb\big]}$. 

For $\delta>1$ there are a total of six linear magnon band crossings ({\it i.e.}, mDPs)  between $\epsilon_{2,3}(\bo)$  located at $\text{D}_2=(\pm k_x^{\text{D}_2}, 0)$ and symmetry related points, where
\begin{align}
k_x^{\text{D}_2}=\arccos\lb\frac{1+2\delta(1-\delta)+f(\delta)}{4\delta^2}\rb.
\end{align}

For $0.5\leq \delta\leq 1$ we obtain magnon band crossings between $\epsilon_{2,3}(\bo)$ along the $k_x=0$ line located at $\text{D}_3=(0,\pm k_y^{\text{D}_3})$, where
\begin{align}
k_y^{\text{D}_3}=\frac{1}{\sqrt{3}}\arccos\lb -1-2\delta+4\delta^2\rb.
\end{align}

The evolution of the magnon band crossing points with varying $\delta$ is depicted in Fig.~\eqref{band}. Evidently, there is no flat band in the strained kagome-lattice ferromagnets for $\delta\neq 1$. The critical point of this system where interesting features emerge is at $\delta=0.5$. As shown in Fig.~\ref{band}(a),  when viewed along the $k_y$ direction, the magnon branches $\epsilon_{1,2}(\bo)$ form linear Dirac magnon cones at $\text{D}_1$, whereas the magnon branches $\epsilon_{2,3}(\bo)$ form quadratic magnon band touching points at  $\text{M}$. However, in Fig.~\ref{band}(b), when viewed along the $k_x$ direction, the magnon branches $\epsilon_{1,2}(\bo)$ form tilted linear Dirac magnon cones at $\text{D}_1$, and the magnon branches $\epsilon_{2,3}(\bo)$ form linear magnon band touching points at  $\text{M}$. Therefore, tilted mDPs and  mSDPs  coexist in the strained kagome-lattice ferromagnetic systems at $\delta=0.5$.  The mSDPs result from the merging of two mDPs  within the magnon branches $\epsilon_{2,3}(\bo)$ for $0.5<\delta<1$. Therefore, the coexisted mDP and mSDP signify a phase transition between two gapless phases in the entire system\footnote{Note that although the mSDPs in the magnon branches $\epsilon_{2,3}(\bo)$ are gap out for $\delta<0.5$, mDPs exist in the magnon branches $\epsilon_{1,2}(\bo)$ at $\text{D}_1$. Therefore, the entire system remains gapless.}. For the tilted mDPs at $\delta=0.5$, the effective Hamiltonian near $\bo=\text{D}_1$ can be written as
\begin{align}
\mathcal H_{\text{sw}}(\vec q+\text{D}_1)=(\epsilon_0+ w q_x){\rm I}_{2\times 2}+v_xq_x\sigma_x+v_yq_y\sigma_y,
\label{ddm}
\end{align}
where  $\sigma_i~(i=x,y,z)$ are Pauli matrices; $\epsilon_0$ is the energy of the mDPs, $v_x$ and $v_y$ are the group velocities along $q_x$ and $q_y$  momentum direction, and $w$ denotes the tilt parameter along the  $q_x$ momentum direction. For the mSDPs  at $\delta=0.5$, the effective Hamiltonian near $\bo=\text{D}_3$ can be written as
\begin{align}
\mathcal{H}_{\text{sw}}(\vec q+\text{D}_3)&=\tilde{\epsilon}_0{\rm I}_{2\times 2}+\tilde v_x\sigma_x q_x+\tilde v_y q_y^2\sigma_y.
\label{sm}
\end{align}

%
%\begin{figure}
%\centering
%\includegraphics[width=1\linewidth]{Gap_M}
%\caption{Color online. The gap between $\epsilon_{2,3}(\bo)$ magnon branches at the $\rm M$-point as a function of $\delta$ for different values of $D/J$.  }
%\label{gap}
%\end{figure} 

\begin{figure}
\centering
\includegraphics[width=1.1\linewidth]{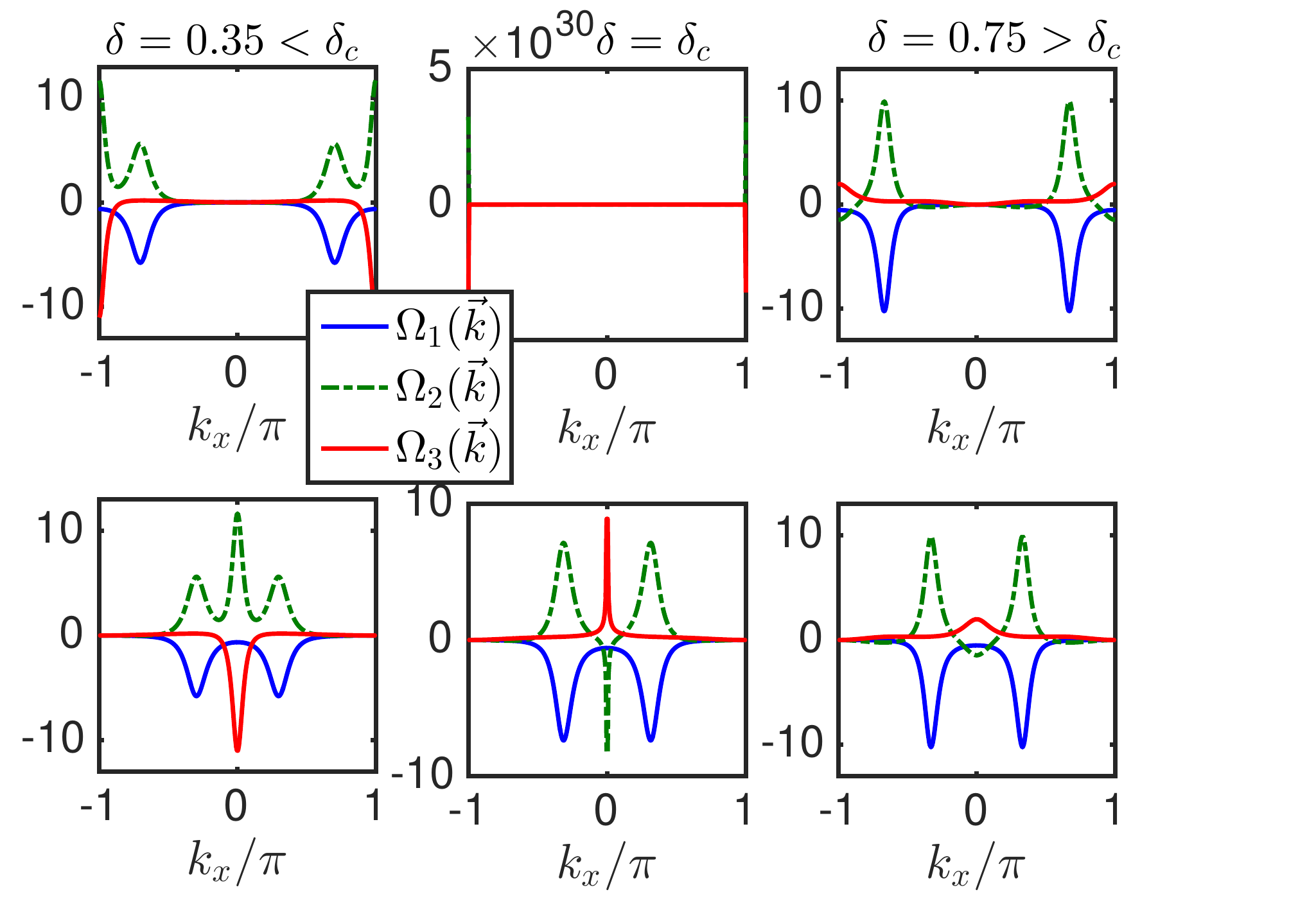}
\caption{Color online. Evolution of the Berry curvature along the $k_x$ direction at $k_y=0$ (top panel) and at $k_y=\pi/\sqrt{3}$, i.e., $\text{K}$--$\text{M}$ line (bottom panel) with $D/J=0.2$. }
\label{bc}
\end{figure}

\section{Strained Kagome Ferromagnets With Dzyaloshinskii-Moriya interaction }
We now consider the effects of the DM interaction due to lack of inversion center on the kagome lattice \cite{alex6}. The DM interaction is given by
\begin{align}
\mathcal H_{\text{DM}}&=\sum_{ \la \ell \ell^\prime\ra } {\vec D}_{\ell \ell^\prime}\cdot{\vec S}_{\ell}\times{\vec S}_{\ell^\prime},
\label{dm}
\end{align}
where ${\vec D}_{\ell \ell^\prime}$ is the DM vector between site $\ell$ and $ \ell^\prime$. In the linear HP spin-boson transformation, only the DM vector parallel to the magnetic field contributes to the noninteracting bosonic Hamiltonian, but other components of the DM vector can be crucial when considering magnon-magnon interactions \cite{cherny}.  The topological aspects of  magnons can be captured clearly in the linear spin wave theory. Therefore, we consider  the out-of-plane DM vector $\vec{D}_{\ell\ell^\prime}=D\hat z$, which is usually present on the kagome lattice. The Fourier transformed DM interaction is given by 
\begin{align}
\Lambda_{\text{DM}}(\vec{k}) &=2it_D
\begin{pmatrix}
0& -\cos k_2& \cos k_3\\
\cos k_2&0&-\cos k_1\\
-\cos k_3&\cos k_1&0
\end{pmatrix},
\label{honn}
\end{align}
 with   $t_D=DS$. Hence $\mathcal H_{\text{sw}}(\vec{k})=\Lambda_0{\rm I}_{3\times 3}-\Lambda(\vec{k})+\Lambda_{\text{DM}}(\vec{k})$.
 
 \subsection{Topological magnon phase transition}
 In Fig.~\eqref{banddm} we show that there is  a topological phase transition at the $\rm{M}$-point between $\epsilon_{2,3}(\bo)$ magnon branches, where the bulk magnon gap closes. It occurs at the critical point 
 \begin{align}  
    \delta_c=\frac{1}{2}\big[ 1-(D/J)^2\big].
    \label{phase}
  \end{align}
  
Indeed, a realistic topological phase transition occurs when  $D< J$. Interesting, this regime of the parameters always exist in realistic kagome materials \cite{alex6,alex6a}.  Thus, we obtain two distinct topological magnon phases for $\delta< \delta_c$ and $\delta>\delta_c$. They have different Chern numbers as will we show in the subsequent subsections. 

Now, we would like to compare  the current results to other phase transitions reported in ferromagnetic systems. First, we compare our result to that of  deformed graphene with next-nearest-neighbour  SOC \cite{lang},  which is equivalent to a distorted honeycomb ferromagnet with a next-nearest-neighbour DM interaction \cite{owe1}. In this  case, the SDPs occur at $\delta_c=2$ as a single entity ({\it i.e} without coexisting with DPs). It is robust against SOC (or DM interaction for magnons), and it transforms to a DP at the topological phase transition point at fixed $\delta_c=2$, separating  topological  $(\delta<2)$ and trivial (for $\delta>2$) insulator \cite{lang}. In contrast, the strained kagome-lattice ferromagnets with DM interaction have no trivial insulator phase, besides the topological phase transition point varies with the DM interaction as shown in Eq.~\eqref{phase}. This  is of interest because different kagome materials have different DM interaction, and should in principle have different phase boundary. 
 
   Second,  let us now compare our result to that of the ideal (isotropic) kagome-lattice ferromagnets with a DM interaction and a second-nearest-neighbour interaction $J_2$ \cite{mp4}, and also the XXZ kagome-lattice ferromagnets with a DM interaction \cite{ran}. In both cases, the topological phase boundaries occur at the $\rm{K}$-point when $D/J=\sqrt{3}\left\vert 2J_2/J-1\right\vert$ and $D/J=\sqrt{3}$  respectively. Unfortunately, these topological phase boundaries cannot be achieved in physical  materials because $D>J$, which is not the case in realistic kagome-lattice materials \cite{alex6,alex6a}.

Indeed, the topological phase transition induced by strain (lattice distortion) in Eq.~\eqref{phase} is evidently different. More importantly,  it is experimentally feasible because a large unrealistic DM interaction is not required.  Therefore, we have established that the value of the DM interaction determines if  a realistic topological phase transition point can be experimentally achieved.  
  \begin{figure}
\centering
\includegraphics[width=1\linewidth]{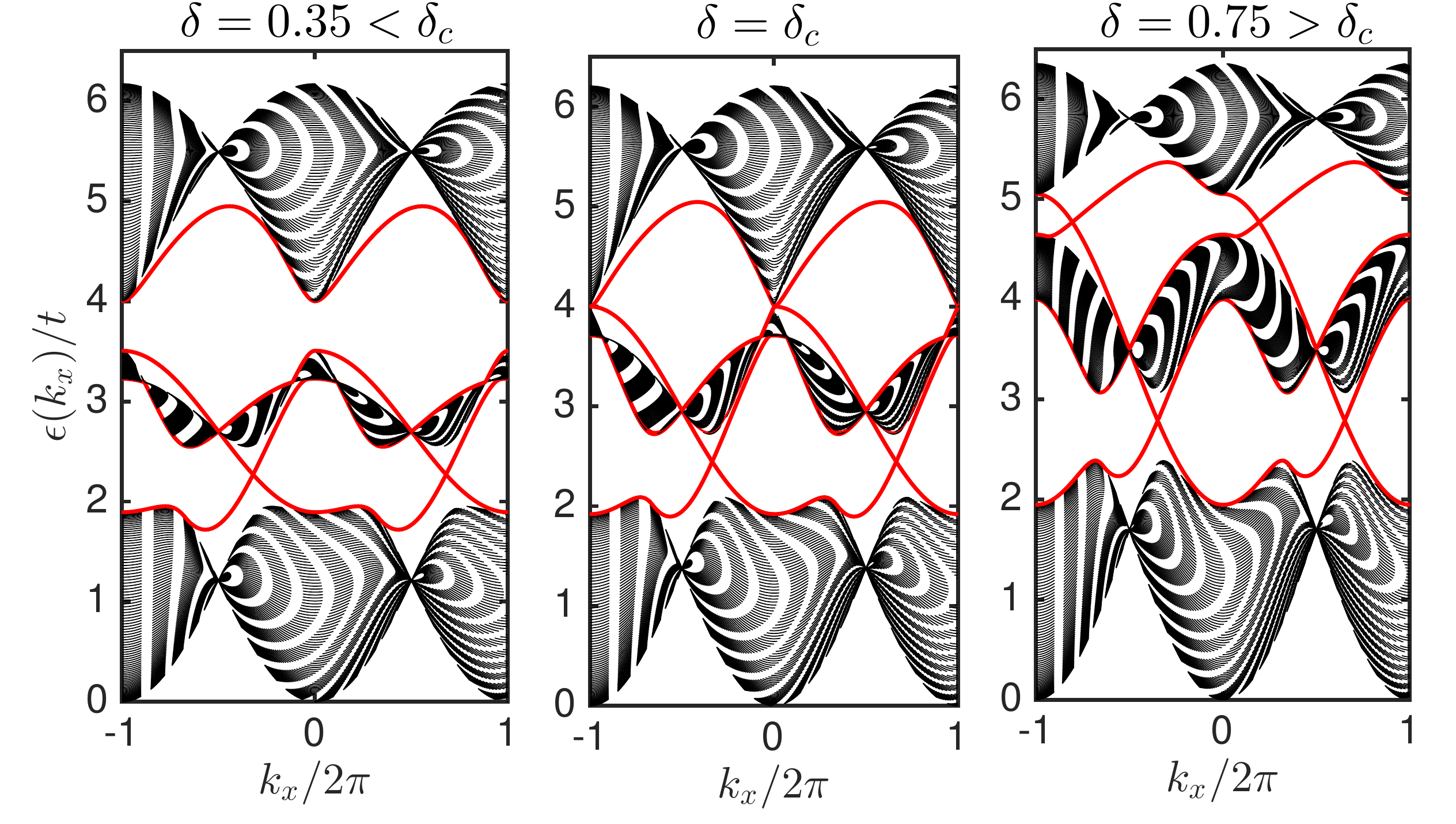}
\caption{Color online. Evolution of the bulk magnon bands (black Zebra lines) and the chiral magnon edge modes (red solid lines ) for a strip geometry with $D/J=0.2$. }
\label{edge}
\end{figure} 
\begin{figure}
\centering
\includegraphics[width=1\linewidth]{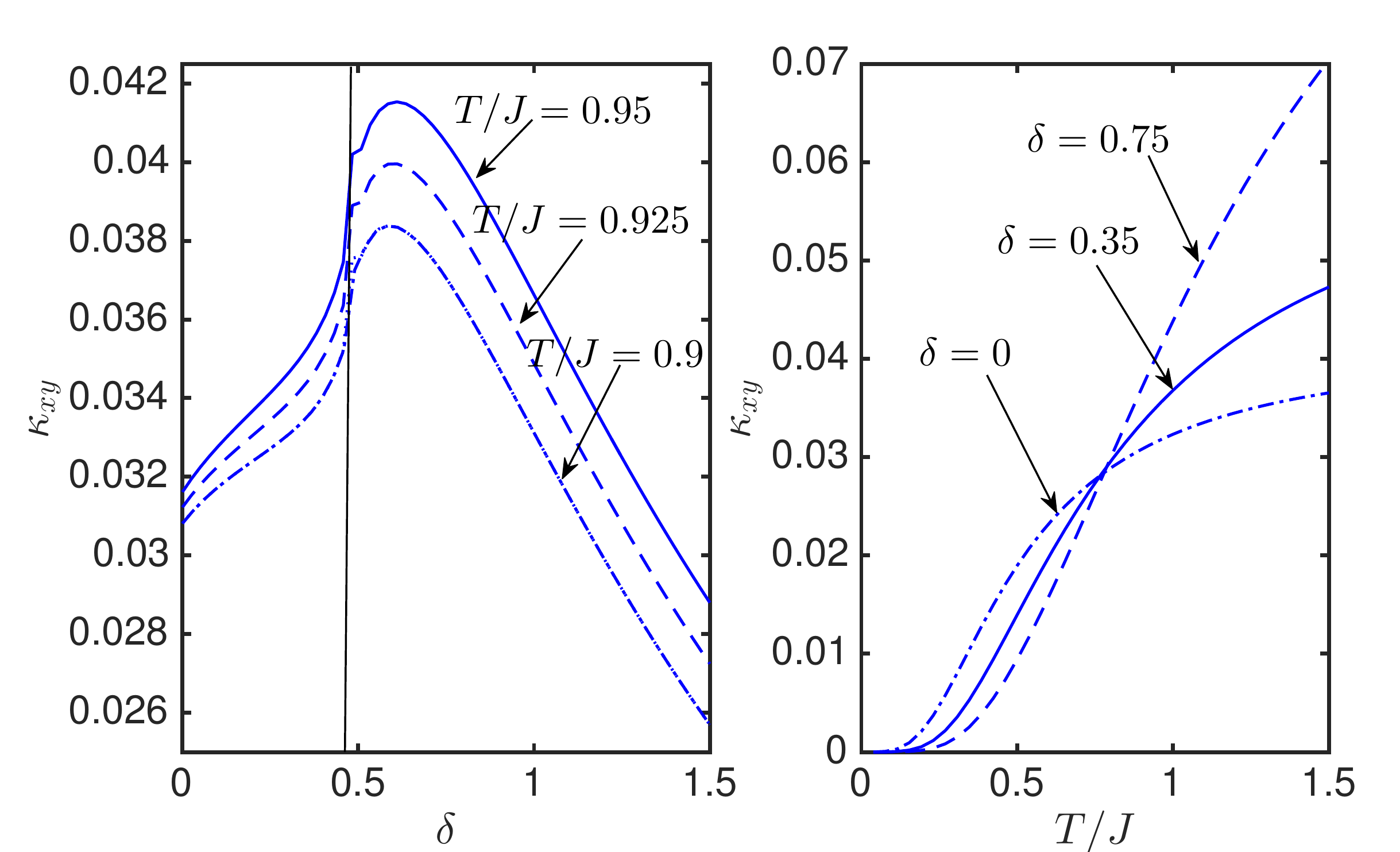}
\caption{Color online. The anomalous thermal Hall conductivity. (Left) $\kappa_{xy}$ vs. $\delta$ for different values of temperature ($T$). The vertical black line indicates the topological phase transition point at $\delta=\delta_c=0.48$ with $D/J=0.2$.  (Right)   $\kappa_{xy}$ vs. $T$ for different values of strain ($\delta$) across the topological phase transition for $D/J=0.2$. The $T$-dependence of  $\kappa_{xy}$  is divergent at $\delta=\delta_c$ (not shown).}
\label{THE}
\end{figure}   
\subsection{Berry curvature}

The presence of the DM interaction and the resulting massive Dirac magnons directly imply that there should be a nonzero Berry curvature.    We now define the Berry curvature of the magnon bands  as
\begin{align}
\Omega_{\alpha}(\vec k)=-\sum_{\alpha^\prime \neq \alpha}\frac{2\text{Im}\big[\braket{u_{\alpha}(\vec k)|\hat v_x|u_{\alpha^\prime}(\vec k)}\braket{u_{\alpha^\prime}(\bo)|\hat v_y|u_{\alpha}(\bo)}\big]}{\big[\epsilon_{ \alpha}(\bo)- \epsilon_{ \alpha^\prime}(\bo)\big]^2},
\label{chern2}
\end{align}
where $\hat v_{x,y}=\partial \mathcal{H}_{\text{sw}}(\vec k)/\partial k_{x,y}$ are the velocity operators,  $u_{\alpha}(\bo)$ are the magnon eigenvectors.  The Berry curvature is nonzero in all regimes of  $\delta$, and  its distribution in momentum space is depicted in Fig.~\eqref{bc}. It shows a very divergent value at the topological magnon phase transition point $\delta_c$. 

\subsection{Chern number}
To describe the topological phase transition we define the associated Chern number  as the integration of the Berry curvature over the BZ,
\begin{align}
\mathcal C_\alpha=\frac{1}{2\pi}\int_{BZ} d^2 \bo~\Omega_{\alpha}(\vec k).
\end{align}

The Chern number changes from  $\mathcal C_\alpha=(-1,1,0)$ in the  regime $\delta< \delta_c$ to $\mathcal C_\alpha=(-1,0,1)$ in the regime $\delta> \delta_c$.  The sign change in the Chern number is consistent with the distribution of the Berry curvature and the redistribution of the magnon bands. Therefore at the critical point  $\delta_c$, the gap between the bulk magnon  bands $\epsilon_{2,3}(\bo)$ close, and the resulting Chern number is ill-defined. This defines a topological phase transition point  separating two distinct mTIs. 

Furthermore,  we  have solved  for the chiral magnon edge modes using a strip geometry with open boundary conditions along the $y$ direction  and infinite along $x$ direction  as depicted in Fig.~\eqref{edge}. A clear feature that emerges is the appearance of the  chiral magnon edge modes in the vicinity of the magnon bulk gaps. Although the chiral magnon edge modes  do not cross each other for $\delta<\delta_c$ in the $\epsilon_{2,3}(\bo)$ magnon branches, the entire system is still a mTI for $\delta<\delta_c$, because the Chern number of the individual magnon band does not vanish identically.

\subsection{Anomalous thermal Hall effect}
One of the crucial implications of nontrivial massive Dirac magnons or mTIs  is that they can transport heat current in the form of the thermal Hall effect \cite{mp1,mp2,mp3,alex6a, thm2, thm3, thm5}. The thermal Hall effect can be understood as a consequence of the Berry curvature induced by the DM interaction.  The general formula for the transverse thermal Hall conductivity $\kappa_{xy}$ has been derived in Ref.~\cite{thm5}. We have used this formula to compute $\kappa_{xy}$ for the current system. The strain and temperature dependence of $\kappa_{xy}$ are shown in Fig.~\ref{THE}.  In the left figure,  $\kappa_{xy}$  changes abruptly at  $\delta=\delta_c$ (vertical black line), an indication of a topological magnon phase transition, which is a direct consequence  of the  Berry curvature in momentum space. In the right figure, we have shown the  $T$-dependence of    $\kappa_{xy}$ with varying $\delta$ across the topological phase boundary. We note that  the  $T$-dependence of $\kappa_{xy}$ is divergent  at  $\delta=\delta_c$ (not shown).

\section{Conclusion}

In summary, we have studied   topological magnon phase transitions  in the two-dimensional strained (distorted) kagome-lattice ferromagnets.  We showed that in the absence of the DM interaction, Dirac and semi-Dirac points coexist in the magnon spectra, and appear between two distinct gapless phases with Dirac points.  The inclusion of the DM interaction gapped out all the magnon Dirac points except at the topological phase transition point where the bulk gap closes between the optical magnon bands, separating two distinct topological magnon insulators with different Chern numbers. In stark contrast to previous studies \cite{mp4,ran}, we have shown that the strained-induced topological magnon phase transitions occur in the realistic regime of the kagome-lattice ferromagnets, {\it i.e.}, $D<J$. 

 We also studied the anomalous thermal Hall effect and showed that it changed abruptly at the critical topological magnon phase transition point. We believe that these results are of great interest in the ongoing experimental studies of Dirac and topological magnons in quantum magnets. Although most kagome ferromagnetic materials may have intrinsic structural distortions, they can also be  tuned experimentally by applying external perturbations such as uniaxial strain or pressure. We propose to apply external strain or  pressure to the currently known magnon topological insulator Cu(1-3, bdc)  \cite{alex6}, whose DM interaction is $D/J=0.15$. Therefore, the current study predicts that a strained-induced topological magnon phase transition should occur at $\delta_c=0.48875$.  In kagome ferromagnetic materials with negligible DM interaction, the current study has predicted a strained-induced  phase transition between two gapless Dirac points at $\delta_c=0.5$, where the phase boundary is defined by the appearance of Dirac points coexisting with semi-Dirac points.     It will also be interesting to investigate the effects of strain in the recently discovered 3D Dirac magnons in Cu$_3$TeO$_6$ \cite{kli, yao,bao}.  As topological magnon phase transition has not been reported experimentally in real magnetic materials, we believe that the current result will  inspire experimentalists to manipulate the intrinsic properties of magnetic materials using external perturbations, which would have a great impact on the future studies of topological magnetic materials.

\section*{ACKNOWLEDGEMENTS}
 Research at Perimeter Institute is supported by the Government of Canada through Industry Canada and by the Province of Ontario through the Ministry of Research and Innovation.


\begin{thebibliography}{99}
\bibitem{cas}
A. H. Castro Neto  et al. Rev. Mod. Phys. {\bf 81}, 109 (2009).
\bibitem{mon1}
Y. Hasegawa et al. Phys. Rev. B {\bf 74}, 033413 (2006).
\bibitem{mon2}
P. Dietl, F. Piechon, and G. Montambaux, Phys. Rev. Lett. {\bf 100}, 236405 (2008).
\bibitem{mon3}
B. Wunsch, F. Guinea, F. Sols, New J. Phys. {\bf 10}, 103027 (2008).
\bibitem{mon5}
V. M. Pereira, A. H. Castro Neto, N. M. R. Peres, Phys. Rev. B {\bf 80}, 045401 (2009).
\bibitem{mon}
G. Montambaux, et al. Phys. Rev. B {\bf 80}, 153412 (2009); Eur. Phys. J. B {\bf 72}, 509 (2009).
\bibitem{mon4}
O. Bahat-Treidel et al., Phys. Rev. Lett. {\bf 104}, 063901 (2010).
\bibitem{mur}
S. Murakami et al. Phys. Rev. B {\bf 76}, 205304
(2007).
\bibitem{lang}
Th. C. Lang et al. Phys. Rev. B {\bf 87}, 205101 (2013).
\bibitem{mag}
J. Fransson, A. M. Black-Schaffer, and A. V. Balatsky, Phys. Rev. B {\bf 94}, 075401 (2016).

\bibitem{owe1}
S. A.  Owerre, J. Phys.: Condens. Matter {\bf 28}, 386001 (2016).
\bibitem{per}
S. S. Pershoguba et al. Phys. Rev. X, {\bf 8} 011010 (2018).
\bibitem{boy}
D. Boyko, A. V. Balatsky, and J. T. Haraldsen, Phys. Rev. B {\bf 97}, 014433 (2018).
\bibitem{yago}
Y. Ferreiros and Mar\'ia A. H. Vozmediano,  Phys. Rev. B {\bf 97}, 054404 (2018).
 \bibitem{dm}
 I. Dzyaloshinsky, J. Phys. Chem. Solids {\bf 4}, 241 (1958).
  \bibitem{dm2}
   T. Moriya, Phys. Rev. {\bf 120}, 91 (1960).
\bibitem{mp1}
H. Katsura, N. Nagaosa, and P. A. Lee, Phys. Rev. Lett. {\bf 104},  066403 (2010).
 \bibitem{mp2} 
  L. Zhang  et al. Phys. Rev. B {\bf 87}, 144101 (2013).
\bibitem{mp3}
 {A.  Mook, J.  Henk, and I. Mertig} {Phys. Rev. B} {\bf 90}, 024412 (2014).
 \bibitem{mp4}
 A.  Mook, J.  Henk, and I. Mertig,  {Phys. Rev. B} {\bf 89}, 134409 (2014).

\bibitem{top1}
 F. D. M. Haldane,    Phys. Rev. Lett. {\bf 61}, 2015  (1988).
 \bibitem{top2}
 C. L. Kane,   and  E. J. Mele,  Phys. Rev. Lett. {\bf 95}, 146802 (2005).
 \bibitem{top3}
  X. -L. Qi  and  S. -C. Zhang,  Rev. Mod. Phys. {\bf 83}, 1057 (2011).
 \bibitem{top4}
 M. Z. Hasan and C. L. Kane,   Rev. Mod. Phys. {\bf 82}, 3045 (2010).
 
 \bibitem{alex6}
R. Chisnell  et al. Phys. Rev. Lett. {\bf 115}, 147201 (2015).

 \bibitem{alex6a}
 Max Hirschberger et al.  \prl {\bf 115}, 106603 (2015).
 

  
\bibitem{bol}
 D. Boldrin et al. Phys. Rev. B {\bf 91}, 220408(R) (2015).
 \bibitem{st1}
 F. Guinea, M. I. Katsnelson, and A. K. Geim, Nat. Phys. {\bf 6}, 30 (2009).
  \bibitem{st2}
 M. Bahramy et al. Nat. Commun. {\bf 3}, 679 (2012).
 
   \bibitem{st3}
 J. Ruan et al. Nat. Commun. {\bf 7}, 11136 (2016).
 \bibitem{st4}
 D. Shao et al. Phys. Rev. B {\bf 96}, 075112 (2017).
  
   \bibitem{fa}
F. Wang, A. Vishwanath, and Y. B. Kim, Phys. Rev. B {\bf 76}, 094421 (2007).



 
\bibitem{cherny}
A. L. Chernyshev,  and  P. A. Maksimov,  Phys. Rev. Lett. {\bf 117}, 187203 (2016).

\bibitem{thm2}
Y. Onose et al.  Science  { \bf 329}, 297 (2010).
\bibitem{thm3}
T. Ideue et al. Phys. Rev. B. {\bf 85}, 134411 (2012).
  \bibitem{thm5}
 R. Matsumoto and S. Murakami, Phys. Rev. Lett. {\bf 106}, 197202 (2011); Phys. Rev. B. {\bf 84}, 184406 (2011).
 
\bibitem{ran}
R. Seshadri and D. Sen, 	Phys. Rev. B {\bf 97}, 134411 (2018). 
   \bibitem{kli}
  K. Li et al.  Phys. Rev. Lett. {\bf 119}, 247202 (2017).
 \bibitem{yao}
  W. Yao   et al.   arXiv:1711.00632 (2017).
\bibitem{bao}
S. Bao  et al. arXiv:1711.02960 (2017).
\end{thebibliography}
\end{document}